\documentclass[10pt]{iopart}

\usepackage{graphicx}
\usepackage{cite}

\begin{document}

\title[Electron impact ionisation of  W$^{\,q+}$ ions for $11\leq q \leq 18$]{Electron-impact single ionisation of W$^\mathbf{q+}$ ions: Experiment and theory for $\mathbf{11\leq q \leq 18}$}

\author{D~Schury$^{1,2}$, A~Borovik, Jr.$^3$, B~Ebinger$^{3,4}$, F~Jin$^{3,5}$, K~Spruck$^1$\footnote{present address: Institut f\"ur Medizinische Physik und Strahlenschutz, THM, 35390 Giessen, Germany}, A~M\"uller$^1$ and S~Schippers$^3$}
\address{$^1$Institut f\"{u}r Atom- und Molek\"{u}lphysik, Justus-Liebig-Universit\"{a}t Gie{\ss}en, Leihgesterner Weg 217, 35392 Giessen, Germany}
\address{$^2$Institut des Nanosciences de Paris, Sorbonne Universit\'{e}, 4 place Jussieu, 75252 Paris, France}
\address{$^3$I. Physikalisches Institut, Justus-Liebig-Universit\"at Gie\ss{}en, Heinrich-Buff-Ring 16, 35392 Giessen, Germany}
\address{$^4$GSI Helmholtzzentrum f\"{u}r Schwerionenforschung GmbH, Planckstr. 1, 64291 Darmstadt, Germany}
\address{$^5$Department of Physics, National University of Defense Technology, Fuyuan Road 1, 410022 Changsha, People's Republic of China}
\ead{daniel.schury@physik.uni-giessen.de, schippers@jlug.de}
\vspace{10pt}
\begin{indented}
\item[]\today
\end{indented}

\begin{abstract}
Absolute cross sections for electron-impact single ionisation (EISI) of multiply charged tungsten ions (W$^{q+}$) with charge states in the range  $ 11 \leq q \leq 18$ in the electron-ion collision energy ranges from below the respective ionisation thresholds up to 1000~eV were measured employing the electron-ion crossed-beams method. In order to extend the results to higher energies, cross section calculations were performed using the subconfiguration-averaged distorted-wave (SCADW) method for electron-ion collision energies up to 150~keV. From the combined experimental and scaled theoretical cross sections rate coefficients were derived which are compared with the ones contained in the ADAS database and which are based on the configuration-averaged distorted wave (CADW) calculations of Loch et al. [Phys. Rev. A \href{https://doi.org/10.1103/PhysRevA.72.052716}{72, 052716 (2005)}]. Significant discrepancies were found at the temperatures where the ions investigated here are expected to form in collisionally ionised plasmas. These discrepancies are attributed to the limitations of the CADW approach and also the more detailed SCADW treatment which do not allow for a sufficiently accurate description of the EISI cross sections particularly at the ionisation thresholds.
\end{abstract}

%
%
\submitto{\JPB}
%
%
%

\section{Introduction}\label{sec:Introduction}

In recent years, collisional and radiative processes involving multiply charged tungsten ions have received much attention by the atomic physics community (see, e.g., \cite{Mueller2015,Loch2005,Kramida2009,Biedermann2009a,Abdallah2011,Ballance2013,Ballance2015,Demura2015b,Clementson2015,Li2015,Kaempfer2016,Li2016a,Harabati2017,Tu2017,Das2018,Krantz2017,Kwon2018,ElMaaref2019,Preval2019,Meyer2019,Jonauskas2019a}), since tungsten will be used as a plasma-facing wall material in future thermonuclear fusion reactors such as ITER. Inevitably, tungsten ions will act as plasma impurities which have to be controlled and understood in order to ensure stable plasma-burning conditions \cite{Philipps2011}. Finding appropriate operating conditions requires extensive modelling of the thermonuclear plasma. To this end, a large amount of reliable data on atomic processes involving all tungsten ionisation stages is needed \cite{Puetterich2010}, including electron-impact ionisation.

Motivated by the importance of tungsten in modern fusion research, a comprehensive experimental programme on studying electron-ion recombination \cite{Schippers2011,Spruck2014a,Badnell2016a}, electron-impact ionisation \cite{Rausch2011,Borovik2016} and photoionisation \cite{Mueller2015c,McLaughlin2016a,Mueller2017a,Mueller2019a} of tungsten ions has been launched \cite{Mueller2015}. As a part of this campaign, the present study addresses electron-impact single ionisation (EISI) of W$^{q+}$ ions with charge states  $11\leq q \leq 18$, thus, extending previous experimental work on EISI of tungsten ions with charge states in the range 1--10  \cite{Stenke1995}. EISI cross sections  for W$^{17+}$ and W$^{19+}$ resulting from this programme have already been published previously \cite{Rausch2011,Borovik2016}. Corresponding rate coefficients have been derived for W$^{19+}$ \cite{Borovik2016} but not for W$^{17+}$. The rate coefficients for EISI of W$^{17+}$ are part of the present results. A comparison of the experimental EISI cross section for W$^{17+}$ with level-to-level distorted wave (LLDW) calculations has been presented by Zhang and Kwon \cite{Zhang2014}.

The main aim of the present study is to benchmark the theoretical work of Loch \etal \cite{Loch2005} who calculated EISI cross sections for all charge states of tungsten ions. These cross sections have been converted into rate coefficients which are available from the ADAS data base \cite{ADAS2017}. Unfortunately, Loch \etal have not published all of the underlying cross sections. In particular the ones that are under scrutiny here are not available. Therefore, EISI cross sections for W$^{q+}$ ions with $11\leq q \leq 18$ have been calculated as part of the present work within the confines of the subconfiguration-averaged distorted wave (SCADW) approach. As described in detail below, the present experimental and theoretical cross sections are used for the derivation of rate coefficients which are eventually compared with the ones available from the ADAS data base.

\section{Experimental setup and procedures}\label{sec:Setup}

Single-ionisation cross sections of W$^{11+}$ up to W$^{18+}$ ions were measured employing the electron-ion crossed-beams method as implemented at the Giessen electron-ion crossed-beams setup \cite{Mueller1985a}. Absolute ionisation cross-sections were obtained by employing the animated-beam technique  at selected electron energies. In addition, to uncover fine details in the ionisation cross section, fine-step energy scans  with an energy-step width of 0.2~eV were performed. Details of the apparatus and of the experimental procedures have already been provided in a previous publications on electron-impact ionisation of W$^{17+}$ ions \cite[and references therein]{Rausch2011}.  Therefore, only a brief description is given here.

\begin{figure}
	\includegraphics[width=\columnwidth]{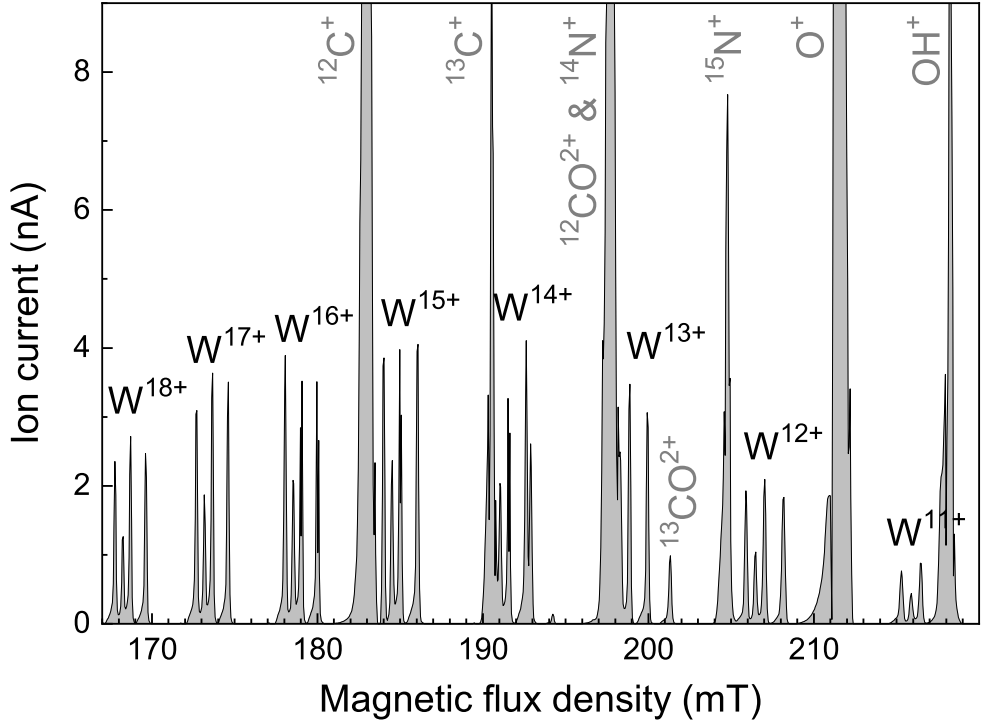}
	\caption{\label{fig:mass_scan} $A/q$ scan of the extracted ion beam, showing tungsten ions in charge states ranging from $q=11$ to $q=18$. The natural abundance pattern of tungsten isotopes can easily be identified  for all the charge states.}
\end{figure}

Tungsten ions were produced by feeding tungsten-hexacarbonyl, W(CO)$_6$, vapor into the plasma chamber of an electron-cyclotron-resonance (ECR) ion source \cite{Schlapp1995a}. In the hot ECR plasma, the W(CO)$_6$ compound dissociates providing tungsten in the gas phase for further ionisation. A mixture of ions produced in the source was extracted and accelerated to an energy of $12q$~keV. The ion beam was mass-over-charge ($A/q$) analyzed by the field of a double focusing 90$^\circ$ dipole magnet. Here, $A$ denotes the atomic mass number. Figure~\ref{fig:mass_scan} shows an $A/q$ scan of the extracted ion beam. Isotope-resolved tungsten-ion species with charge states between  $q=11 $ and $q=18$ can be discerned. The natural tungsten isotope abundance pattern can easily be identified for all charge states. For the ionisation measurements the isotopes $^{184}$W or $^{186}$W, depending on the charge state, were used, which were well enough separated from any parasitic ion species such as C$^+$, N$^+$, CO$^{2+}$, O$^+$ or OH$^{+}$ (cf.~figure~\ref{fig:mass_scan}) for all measured charge states.

After an additional charge-state purification by an electrostatic spherical deflector, the primary ion beam was collimated by two rectangular apertures positioned 195~mm apart from each other. The beam sizes were in the ranges $0.6\times0.6-1.2\times1.2$~mm$^2$ for the absolute cross-section measurements and $1.5\times1.5-2\times2$~mm$^2$ for the energy-scan measurements. Under these conditions, the collection efficiency for both the primary ion beam and the product ions produced in the interaction region was nearly 100\%. In the interaction region, the ion beam was crossed by a ribbon-shaped electron beam produced by the high-power electron gun described by Becker \textit{et al.} \cite{Becker1985}. The gun delivers electron currents of up to 450~mA at a maximum electron energy of 1000~eV. Product and primary ions were separated from one another by a second dipole magnet, identical to the first one. The field of this magnet was set such that the desired product ions were deflected by 90$^\circ$ to enter the single-particle detector \cite{Rinn1982} down-beam behind a 180$^\circ$ out-of-plane deflector which suppresses background arising from stray particles and photons. The primary ions were detected by a Faraday cup appropriately positioned inside the vacuum chamber of the second magnet.

The systematic uncertainties of the absolute cross-section measurements were determined as the quadrature sum of the uncertainties of the parameters entering the cross-section calculation, resulting in a total uncertainty of 6.5\% \cite{Rausch2011} except for W$^{13+}$ where additional uncertainties due to background subtraction result in a systematic error of 16\%.  An additional energy-dependent systematic uncertainty arises from the transmission of the electron beam through the interaction region. This is less than 1\% at energies greater than 120 eV and thus does not significantly contribute to the error budget of the present measurements.

The statistical uncertainty of the absolute cross-section measurements was usually kept at 2.5\% or better. Only for values very close to the threshold, such high statistical accuracy could not be achieved within a reasonable measurement time because of small cross sections at those energies combined with lower electron currents available at low electron energies. This results in low event rates which compete with background arising from electron stripping by the fast parent ions in the residual gas. The resulting total experimental uncertainty for the absolute cross-section values well beyond the ionisation threshold is 7\% (17\% for W$^{13+}$) or lower. For the energy-scan measurements, the statistical error at energies well above the ionisation onset is 1.5\% or lower. The uncertainty of the electron-ion collision energy-scale amounts to 0.3\%.

\section{Theoretical method}\label{sec:theory}

For the calculation of the EISI cross sections for W$^{q+}$ ions with $11\leq q \leq 18$, the SCADW approach was used which is based on relativistic wave functions and relativistic atomic configurations. In contrast, the CADW method is entirely nonrelativistic. In the present SCADW calculations, the ionisation cross section is represented as a sum of the cross sections for the \emph{direct ionisation} (DI) and indirect \emph{excitation-autoionisation} (EA) processes, i.e.,
\begin{equation}
\sigma = \sum_{ij}\sigma_{ij}^{\mathrm{DI}} + \sum_{ij}\sum_k\sigma_{ik}^{\mathrm{CE}}B_{kj}
\end{equation}
with the direct-ionisation cross section $\sigma_{ij}^{\mathrm{DI}}$ of an initial configuration $i$ to a final configuration $j$ and the cross section $\sigma_{ik}^{\mathrm{CE}}$ for collisional excitation of an initial configuration $i$ to an autoionising configuration $k$. The latter was multiplied by the corresponding branching ratio $B_{kj}$ for autoionisation. This ratio represents the probability for an excited level to decay via an Auger process rather than by photoemission.  All atomic quantities were calculated with the Flexible Atomic Code (FAC) \cite{Gu2008}. Subsequently, the cross sections were assembled from these quantities using a newly developed code.

\begin{table}
\caption{\label{tab:gs}Ground configurations with their numbers of fine-structure levels and associated  ionisation energies of W$^{q+}$ ions according to the present SCADW calculations. For comparison, the ionisation energies from the NIST atomic spectra data base \cite{Kramida2009,ASD2019} are also provided. It is noted that, in this data base, the [Kr] $4d^{10}\,4f^{14}\,5s^2$ configuration is found to be the ground configuration of W$^{12+}$.}
\begin{indented}
	\item[]\begin{tabular}{clrcc}\br
		 & Ground & No. of &\multicolumn{2}{c}{Ionisation Energy (eV)} \\
		$q$ & Configuration & Levels & SCADW & NIST  \\
        \mr
		11 & [Kr] $4d^{10}\,4f^{13}\,5s^2\,5p^2$  & 30&  $231.5\pm3$ & $231.6\pm1.2$  \\
		12 & [Kr] $4d^{10}\,4f^{13}\,5s^2\,5p^1$  & 12& $254.6\pm3$ & $258.3\pm1.2$  \\
		13 & [Kr] $4d^{10}\,4f^{13}\,5s^2$ &  2 & $289.0\pm3$ & $290.7\pm1.2$  \\
		14 & [Kr] $4d^{10}\,4f^{12}\,5s^2$ & 13 &$324.8\pm3$ & $325.3\pm1.5$  \\
		15 & [Kr] $4d^{10}\,4f^{11}\,5s^2$ & 41 &$360.1\pm3$ & $361.9\pm1.5$  \\
		16 & [Kr] $4d^{10}\,4f^{11}\,5s^1$  & 82 &$385.5\pm3$ & $387.9\pm1.2$  \\
		17 & [Kr] $4d^{10}\,4f^{11}$  & 41&$419.3\pm3$ & $420.7\pm1.4$  \\
		18 & [Kr] $4d^{10}\,4f^{10}$  & 107 &$459.1\pm3$ & $462.1\pm1.4$  \\
		\br			
	\end{tabular}
\end{indented}
\end{table}

\begin{table}
\caption{\label{tab:di} Overview over all atomic subshells considered in the calculations of DI cross sections. $\bullet$: present SCADW calculations, $\circ$: CADW calculations of Loch \etal \cite{Loch2005}.}
\begin{indented}
	\item[]	\begin{tabular}{ccccccc}\br
		$q$ & $4s$ & $4p$ & $4d$ & $4f$ & $5s$ & $5p$  \\
\mr
		11 &  & $\bullet\phantom{\circ}$ & $\bullet\circ$ & $\bullet\circ$ & $\bullet\circ$ & $\bullet\circ$ \\
		12 &  & $\bullet\phantom{\circ}$ & $\bullet\circ$ & $\bullet\circ$ & $\bullet\circ$ & $\bullet\circ$ \\
		13 &  & $\bullet\phantom{\circ}$ & $\bullet\circ$ & $\bullet\circ$ & $\bullet\circ$ &  \\
		14 &  & $\bullet\phantom{\circ}$ & $\bullet\circ$ & $\bullet\circ$ & $\bullet\circ$ &  \\
		15 & $\bullet\phantom{\circ}$ & $\bullet\phantom{\circ}$ & $\bullet\circ$ & $\bullet\circ$ & $\bullet\circ$ &  \\
		16 &  & $\bullet\phantom{\circ}$ & $\bullet\circ$ & $\bullet\circ$ & $\bullet\circ$ &  \\
		17 &  & $\bullet\circ$ & $\bullet\circ$ & $\bullet\circ$ &  &  \\
		18 &  & $\bullet\circ$ & $\bullet\circ$ & $\bullet\circ$ &  &  \\
		\br
	\end{tabular}
\end{indented}
\end{table}

\begin{table}
\caption{\label{tab:ea}Summary of the EA channels that were taken into account in the present calculations. For every charge state and initial shell the maximum quantum numbers $n$ and $l$ are given for up to which EA cross sections were calculated.}
\begin{indented}
	\item[]	\begin{tabular}{cc@{\extracolsep{2mm}}cc@{\extracolsep{2mm}}cc@{\extracolsep{2mm}}cc@{\extracolsep{2mm}}cc@{\extracolsep{2mm}}c}\br
		& \multicolumn{2}{c}{5\textit{s}} & \multicolumn{2}{c}{4\textit{f}} & \multicolumn{2}{c}{4\textit{d}} & \multicolumn{2}{c}{4\textit{p}} & \multicolumn{2}{c}{4\textit{s}} \\
		$q$ & $n$ & $l$ & $n$ & $l$ & $n$ & $l$ & $n$ & $l$ & $n$ & $l$  \\
\mr
		11 & 25 & 8 & 25 & 8 & 25 & 8 & 5 & 8 & &  \\
		12 & 25 & 8 & 25 & 8 & 25 & 8 & 5 & 4 & & \\
		13 & 25 & 8 & 25 & 8 & 25 & 8 & 9 & 8 & \multicolumn{2}{c}{4f, 5p}  \\
		14 & & & & & 25 & 8 & 9 & 8 & \multicolumn{2}{c}{4f 5p 5d}  \\
		15 & & & 25 & 8 & 25 & 8 & 9 & 8 & \multicolumn{2}{c}{4f 5p-5d 6s}  \\
		16 & & & 25 & 8 & 25 & 8 & 9 & 8 & \multicolumn{2}{c}{5s-5g 6s-6d}  \\
		17 & & & & & 25 & 8 & 21 & 8 & 6 & 5  \\
		18 & & & & & 25 & 8 & 25 & 8 & 9 & 8  \\
		\br
	\end{tabular}
\end{indented}
\end{table}

Table~\ref{tab:gs} lists the ground configurations of W$^{11+}$ up to W$^{18+}$ as well as the presently calculated respective ionisation potentials which, within the mutual uncertainties, agree with the values from the NIST atomic spectra database \cite{Kramida2009,ASD2019}. The only difference concerns the designation of the W$^{12+}$ ground configuration. The present theoretical uncertainties in table \ref{tab:gs} were estimated by comparison of the SCADW results with separately performed LLDW calculations.

Tables~\ref{tab:di} and \ref{tab:ea} provide an overview over the $nl$ subshells and the excitation channels that were considered in the present calculations of DI and EA cross sections, respectively. Excitations with a threshold above the double-ionisation threshold were left out in the EA calculations, since these states are assumed to completely decay via two or more sequential autoionisation processes. For W$^{14+}$, the excitation of a $4d$ electron straddles the ionisation threshold. Such a situation cannot be treated well in the SCADW approach. Comparisons with more elaborate LLDW calculations \cite{Jin2019} suggest that the SCADW calculations underestimate the respective EA cross section by about $0.6\times10^{-18}$~cm$^2$ for energies close to the excitation threshold. Likewise, a difference of $0.26\times10^{-18}$~cm$^2$ was found for EA involving the $4d$ shell of W$^{15+}$. For W$^{18+}$, the present SCADW $4d\to5f$ and $4d\to6p$ excitation energies are below the respective ionisation thresholds. However, the more exact LLDW calculations suggests that these excitation energies are above the ionisation thresholds. Nevertheless, these excitations were not included into the present SCADW calculations of the W$^{18+}$ EA cross sections for the sake of employing the SCADW method consistently for all charge states investigated here.

\section{Results}\label{sec:comparison}

\subsection{Cross sections}

\begin{figure}
\includegraphics[width=0.97\columnwidth]{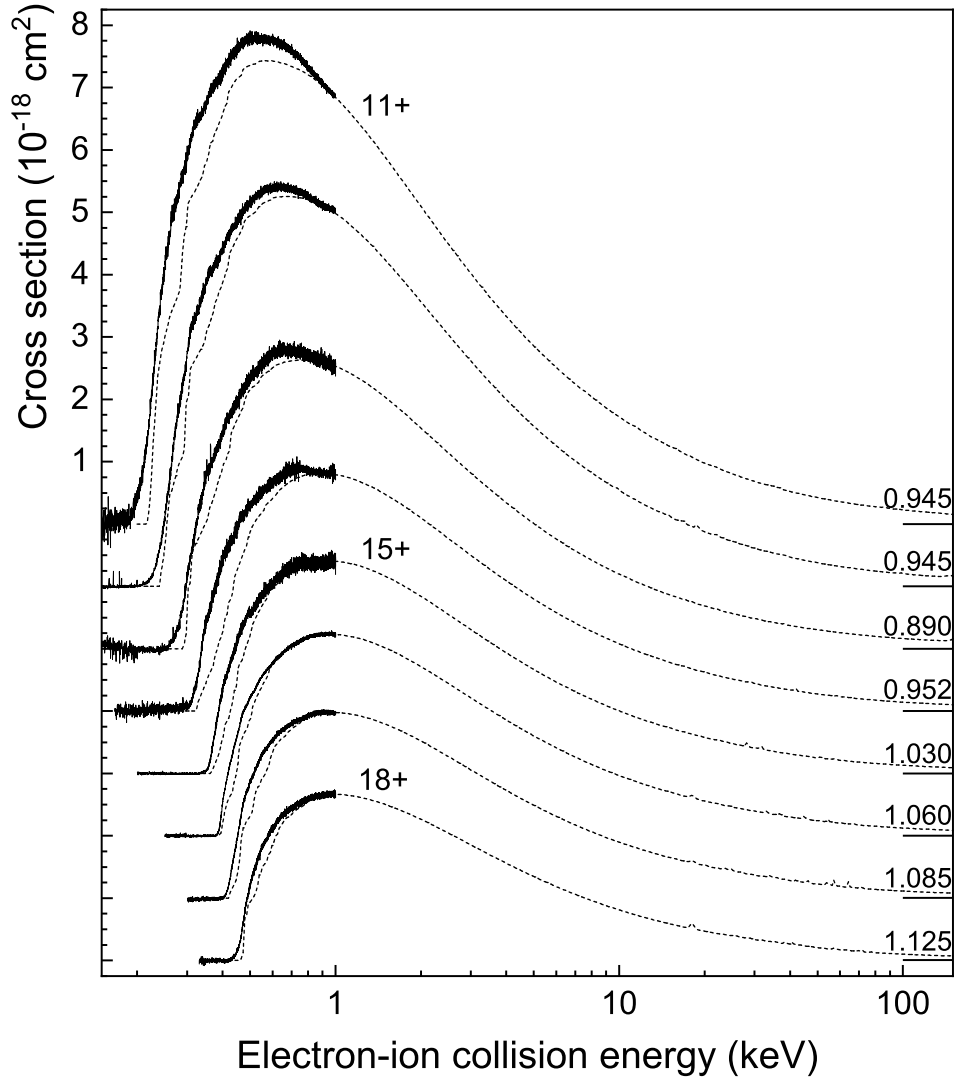}
\caption{\label{fig:Overview}Overview of the presently measured and calculated EISI cross sections of W$^{q+}$ ions with $11\leq q \leq 18$. The black solid lines represent the normalised fine-step energy scan data. The short-dashed lines are the present SCADW cross sections scaled by a constant factor to match the experimental data at the highest experimental electron-ion collision energies. The individual scaling factors are given as labels of the corresponding theoretical cross-section curves. For better visibility of the individual cross sections they have been shifted by $-(q-11)\times10^{-18}$~cm$^{2}$ on the cross section scale. The small excursions of the theoretical cross sections at energies above 10.000 eV are due to slight numerical instabilities. The experimental cross sections for W$^{17+}$ were already published previously \cite{Rausch2011}.}
\end{figure}

\begin{figure}
	\includegraphics[width=0.97\columnwidth]{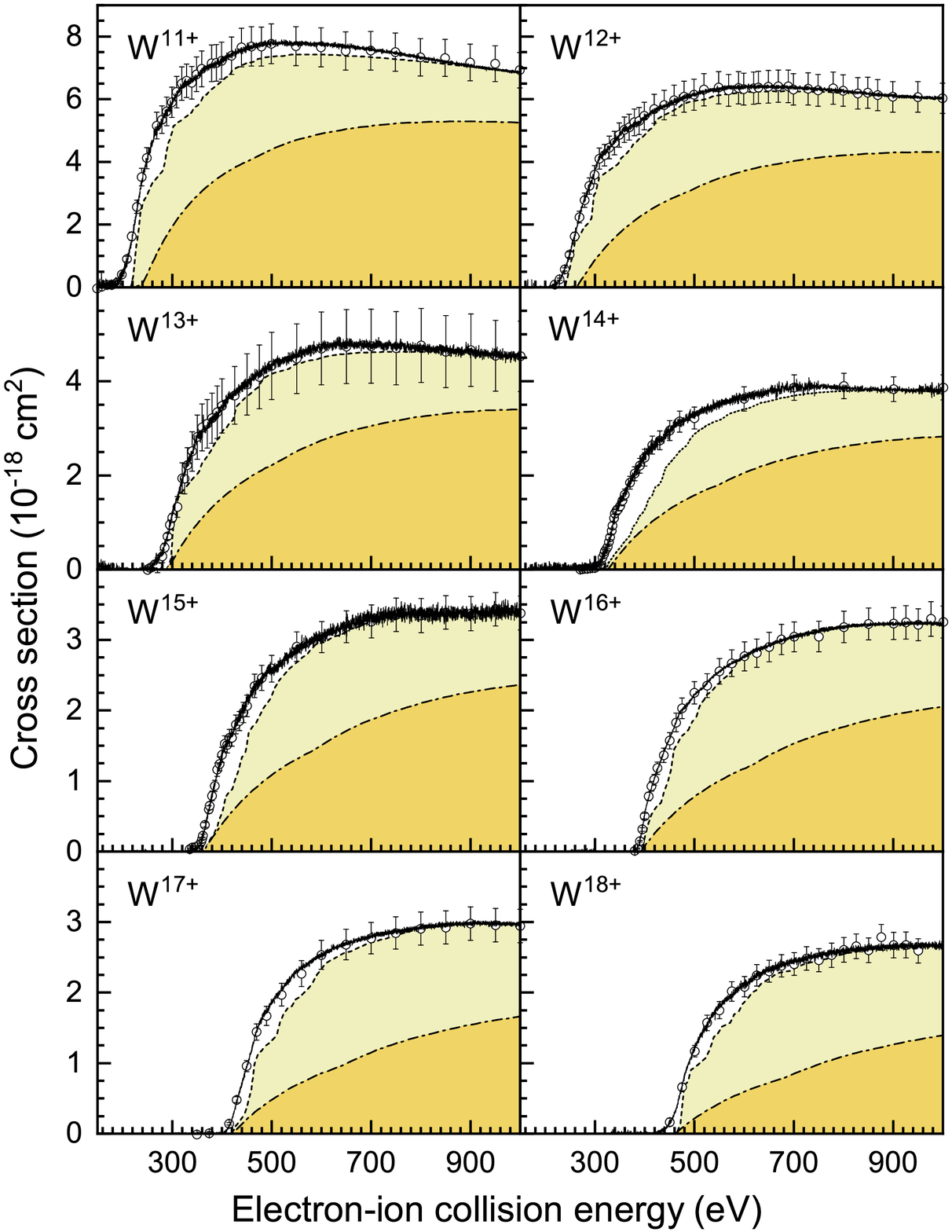}
	\caption{Experimental (full lines and open symbols) and scaled theoretical (shaded curves) cross sections for electron-impact ionisation of W$^{q+}$ ions with $11\leq q \leq 18$. The open symbols are the absolute experimental data points. The associated error bars comprise statistical and systematic uncertainties. The dash-dotted curves are the scaled theoretical cross sections for DI. The short-dashed curves comprise DI and EA contributions. The same scaling factors were used for the theoretical cross sections as in Fig.~\ref{fig:Overview}. The experimental cross sections for W$^{17+}$ were already published previously \cite{Rausch2011}.}\label{fig:Overview2}
\end{figure}

Figure~\ref{fig:Overview} shows all  measured and calculated EISI cross sections for W$^{q+}$ ions with $11\leq q \leq 18$. The displayed experimental data are the results of the energy-scan measurements which were normalised to the individually measured absolute data points. The latter are not shown for the sake of clarity, but are displayed in figure \ref{fig:Overview2}. The theoretical cross sections were scaled to the experimental ones at the highest experimental energies. This scaling is required for the derivation of plasma rate coefficients as explained below. The energy-independent scaling factors are displayed as labels of the corresponding cross-section curves. For W$^{13+}$, W$^{17+}$, and W$^{18+}$ the scaling factors differ from one by more than the uncertainty of the experimental cross-section scale. This is largely attributed to the limited accuracy of the present theoretical SCADW approach which bears rather large uncertainties, in particular for the DI cross sections.

All cross sections exhibit a rather smooth dependence on the electron-ion collision energy without any prominent resonance or step features. The cross-section maxima are found to be within the experimental energy range for W$^{11+}$ to W$^{14+}$ with peak values of $7.9\times10^{-18}$~cm$^2$, $7.5\times10^{-18}$~cm$^2$, $5.0\times10^{-18}$~cm$^2$, and $4.1\times10^{-18}$~cm$^2$, respectively. Starting with W$^{15+}$, the maximum moves out of the energy range of the experiment. This imposes an additional uncertainty on the scaling of the theoretical cross sections to the experimental ones for  W$^{15+}$ to W$^{18+}$. It is noted that the corresponding scaling factors exhibit a monotonic increase with increasing charge state which is attributed to the fact that the scaling is applied at energies below the cross section maxima. It must be concluded that the uncertainties of the scaled theoretical cross sections are much larger than the $\pm$7\% uncertainty of the experimental cross sections. On the basis of the scaling factors displayed in figure \ref{fig:Overview}, an uncertainty of $\pm$20\% is conservatively assigned to all scaled theoretical cross sections. This also provides an estimate of the uncertainty of the SCADW calculations.

Figure \ref{fig:Overview2} displays the same data as figure \ref{fig:Overview}, but, over a restricted energy range to provide a better view of the threshold regions. In addition, the absolute experimental data points are displayed and the DI and EA contributions to the theoretical cross sections are plotted separately. Obviously, the scaled theoretical SCADW cross sections underestimate the experimental cross sections at energies close to the ionisation thresholds. The discrepancies decrease with increasing charge state which is partly due to the scaling at energies below the cross-section maxima. The present calculations reveal that EA processes contribute significantly to the total EISI cross sections for all charge states under investigation.

For all charge states except W$^{15+}$ the experimental EISI cross sections have an onset at lower energies than the ground-configuration ionisation potentials from table~\ref{tab:gs}. This indicates the presence of long lived metastable excited levels in the primary beams.  Possible candidates for such levels are fine-structure excited levels of the ground configurations and also levels of excited configurations with the same parity as the respective ground configuration, as has already been discussed, e.g., for W$^{17+}$ \cite{Rausch2011} and W$^{19+}$ \cite{Borovik2016}. The fractional populations of the metastable levels are generally not known. Common assumptions are a statistical level distribution or a Maxwell-Boltzmann distribution which accounts for the plasma temperature in the ion source \cite{Jonauskas2019a}.
In principle, more detailed information about the metastable fractions can be obtained from comparisons with theoretical cross sections if these are known with sufficient precision. This has been shown to yield consistent results for few-electron ions \cite{Borovik2009}. For many-electron ions, however, the computational effort is much larger \cite{Jonauskas2019a} and beyond the scope of the present work.

The present SCADW approach assumes a statistical population of the fine-structure levels within the ground configuration (cf.~table~\ref{tab:gs}). This approach leads to deviations between SCADW and experimental cross sections in the threshold region (figure~\ref{fig:Overview2}) because the SCADW calculations are based on the average ionisation energy of each subconfiguration instead of the appropriate individual ionisation thresholds of all the contributing levels. These excited levels have lower ionisation thresholds. Disregard of this energy splitting is also partly responsible for the underestimation of the cross sections near the threshold. This can be remedied by much more detailed LLDW calculations as will be shown for W$^{14+}$ as an example in a follow-up publication \cite{Jin2019}. In any case, the relatively small deviations from unity of the scaling factors for the ground-configuration cross-sections (figure \ref{fig:Overview}) hint to an only moderate influence of possible metastable primary ions on the measured cross sections.

\subsection{Plasma rate coefficients}\label{sec:pra}

In order to be able to compare with the EISI data from the ADAS data base \cite{ADAS2017}  plasma rate coefficients (PRC) for ionisation of W$^{11+}$ to W$^{18+}$ ions were derived from the present experimental cross sections using the procedures employed already for W$^{19+}$ \cite{Borovik2016}. Accordingly, the experimental cross section is multiplied by the relative electron-ion velocity and the resulting product is integrated over a Maxwellian velocity distribution \cite{Mueller1999}.

\begin{figure}
	\includegraphics[width=\columnwidth]{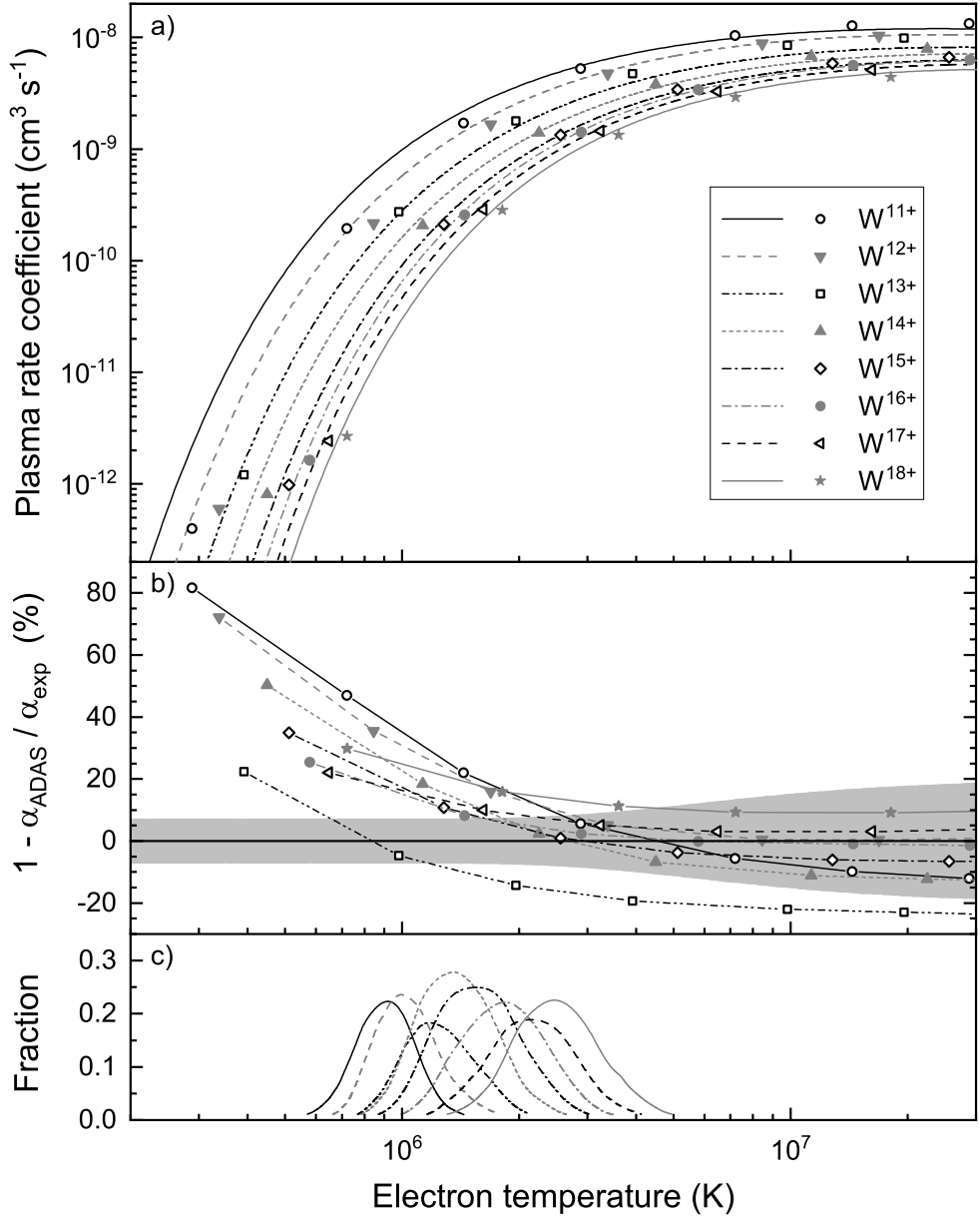}
	\caption{\label{fig:PRC} a) Present experimentally-derived plasma rate coefficients $\alpha_\mathrm{exp}$ for electron-impact single-ionisation of W$^{11+}$ to W$^{18+}$ ions (smooth lines as explained in the inset). The symbols represent the rate coefficients $\alpha_\mathrm{ADAS}$ from the ADAS data base \cite{ADAS2017}. b) Relative deviations $1-\alpha_\mathrm{ADAS}/\alpha_\mathrm{exp}$ of the ADAS rate coefficients from the present results. The symbols appear at the energies for which values of the ADAS rate coefficients are available. The lines are drawn to guide the eye. The grey area marks the uncertainty of the present experimentally-derived rate coefficients stemming from the $\pm7\%$ uncertainty of the experimental cross sections ($\pm17\%$  for W$^{13+}$) and the $\pm20\%$ uncertainty of the scaled theoretical cross section. c) Charge-state fractions from charge balance calculations \cite{Preval2019} providing a coarse indication for the relevant temperature range.}
\end{figure}

For the derivation of a PRC for electron temperatures $T_e$ up to $T_e^{(\mathrm{max})}$, the underlying cross section must be known from the ionisation threshold $I_0$ up to an energy $E^{(\mathrm{max})} = 6 k_B T_e^{(\mathrm{max})} + I_0$ with $k_B$ denoting the Boltzmann constant \cite{Fogle2008}. Previous charge-balance calculations suggest that, in collisionally ionised plasmas, tungsten ions with charge states 11--18 form at temperatures ranging approximately from $5\times10^5$ to $5\times 10^6$~K \cite{Preval2019} (Fig.~\ref{fig:PRC}c). These temperatures probably  bear large uncertainties since the calculations are based on untested theoretical cross sections. Nevertheless, they allow for a meaningful choice of an upper temperature limit at $T_e^{(\mathrm{max})} = 2\times10^7$~K which corresponds to $E^{(\mathrm{max})} \approx 11$~keV. This is much higher than the 1 keV upper limit of the experimental electron-energy scale. Therefore, the present experimental cross sections were extrapolated by our SCADW results (actually to a maximum energy of 150~keV). To match the theoretical cross sections to the experimental ones at the highest experimental electron energies, the theoretical cross sections were multiplied by the energy-independent factors given in Fig.~\ref{fig:Overview}.

Figure \ref{fig:PRC}a shows the resulting PRCs in comparison with the ones available from the ADAS database \cite{ADAS2017}. There are significant differences in particular at the temperatures where the tungsten charge states 11--18 form in collisionally ionised plasmas. A coarse estimate for these temperatures can be obtained from Fig.~\ref{fig:PRC}c. The relative deviations of the ADAS rate coefficients from the present experimentally derived results amounts to up to 80\% for W$^{11+}$ (Fig.~\ref{fig:PRC}b). The deviations are larger for the lower charged ions than for the higher charged ones, i.e., the deviations increase with the complexity of the atomic structure (cf.~Tab.~\ref{tab:gs}). The fact that the deviations tend to decrease with increasing temperature suggests that the cross sections from Loch \etal \cite{Loch2005}, which were used for deriving the ADAS rate coefficients, underestimate the near-threshold EISI cross sections for the tungsten ions considered here. Unfortunately, the cross-section data of Loch \etal are \emph{not} available for a closer comparison.

\begin{table}
	\caption{\label{tab:PRC}Parameters $I_0$ in eV and $k_i$ in eV$^{3/2}$~m$^3$~s$^{-1}$ for expressing the present plasma rate coefficients via Eqs.~\ref{eq:scaled_temp}--\ref{eq:alphas} in the temperature ranges
$2\times10^5 - 3\times10^7$~K  for W$^{11+}$--W$^{17+}$ and $4\times10^5 - 8\times10^7$~K for W$^{18+}$. In these temperature ranges the parameterised rate coefficients deviate by less than 2\% from the experimentally derived curves. The numbers in square brackets denote powers of 10, i.e., $[x]$ corresponds to a multiplication of the preceding number by $10^x$.}
\begin{indented}
	\item[]	\begin{tabular}{crrrr}
\br
		        & W$^{11+}$                & W$^{12+}$              & W$^{13+}$               & W$^{14+}$ \\
\mr
		$ I_0 $ &  1.89400[\phantom{-}2]   &  2.01800[\phantom{-}2] &   2.56800[\phantom{-}2] &  2.91200[\phantom{-}2] \\
		$ k_0 $ &  1.62869[-5]             & -4.30311[-6]           &   7.78720[-6]           &  3.83113[-6]           \\
		$ k_1 $ &  2.92628[-4]             &  2.28614[-4]           &   6.31766[-4]           &  9.10962[-4]           \\
		$ k_2 $ & -1.25072[-4]             &  1.56889[-3]           &  -4.00373[-3]           & -7.41404[-3]           \\
		$ k_3 $ & -1.19602[-2]             & -2.30676[-2]           &   1.28178[-2]           &  3.60079[-2]           \\
		$ k_4 $ &  8.30703[-2]             &  1.24526[-1]           &  -9.26167[-3]           & -1.09027[-1]           \\
		$ k_5 $ & -2.77200[-1]             & -3.78357[-1]           &  -7.47009[-2]           &  2.05145[-1]           \\
		$ k_6 $ &  5.31098[-1]             &  6.98097[-1]           &   2.81317[-1]           & -2.26890[-1]           \\
		$ k_7 $ & -5.95049[-1]             & -7.75589[-1]           &  -4.51015[-1]           &  1.23438[-1]           \\
        $ k_8 $ &  3.62956[-1]             &  4.77561[-1]           &   3.57553[-1]           & -9.18026[-3]           \\
		$ k_9 $ & -9.32210[-2]             & -1.25293[-1]           &  -1.14328[-1]           & -1.34376[-2]           \\
\br
		        & W$^{15+}$                & W$^{16+}$              & W$^{17+}$               &  W$^{18+}$\\
\mr
		$ I_0 $ &  3.39000[\phantom{-}2]   &  3.81500[\phantom{-}2] &  4.03900[\phantom{-}2]  &  4.28100[\phantom{-}2]         \\
		$ k_0 $ & -4.79355[-6]             &  9.80416[-5]           &  4.02252[-5]            &  9.75816[-6]          \\
		$ k_1 $ &  1.97326[-3]             &  4.86978[-4]           &  2.26975[-3]            &  2.51270[-3]         \\
		$ k_2 $ & -2.37255[-2]             & -8.18620[-3]           & -3.16756[-2]            & -2.92481[-2]        \\
		$ k_3 $ &  1.68014[-1]             &  5.97534[-2]           &  2.39946[-1]            &  1.89761[-1]         \\
		$ k_4 $ & -7.50995[-1]             & -2.53577[-1]           & -1.11620[\phantom{-}0]  & -7.62806[-1]        \\
		$ k_5 $ &  2.16681[\phantom{-}0]   &  6.71410[-1]           &  3.30780[\phantom{-}0]  &  1.96322[\phantom{-}0]      \\
		$ k_6 $ & -4.01635[\phantom{-}0]   & -1.12299[\phantom{-}0] & -6.25023[\phantom{-}0]  & -3.23398[\phantom{-}0]     \\
		$ k_7 $ &  4.61065[\phantom{-}0]   &  1.15248[\phantom{-}0] &  7.27957[\phantom{-}0]  &  3.29420[\phantom{-}0]      \\
		$ k_8 $ & -2.97967[\phantom{-}0]   & -6.62196[-1]           & -4.75776[\phantom{-}0]  & -1.88827[\phantom{-}0]     \\
		$ k_9 $ &  8.27992[-1]             &  1.63023[-1]           &  1.33406[\phantom{-}0]  &  4.65500[-1]       \\
\br
	\end{tabular}
\end{indented}
\end{table}

In order to provide a simple expression for plasma modeling purposes, the present PRCs are expressed in terms of the Burgess-Tully model which introduces a scaled electron temperature
\begin{equation}
\chi = 1- \frac{\ln\left(2\right)}{\ln\left(\frac{k_B T_e}{I_0}+2\right)}\label{eq:scaled_temp}
\end{equation}
and a scaled rate coefficient $ \rho\left(\chi\right) $ \cite{Dere2007}. The PRC $\alpha(T_e)$ can then be obtained as
\begin{equation}
\alpha(T_e) = \frac{E_1\left(\frac{I_0}{k_B T_e}\right)\rho\left(\chi\right)}{I_0 \sqrt{k_B T_e}}\label{eq:prc}
\end{equation}
where $ E_1 $ is the first exponential integral, $ k_B $ is the Boltzmann constant, and $ I_0 $ represents the ionisation threshold. The scaled rate coefficient is finally approximated by a polynomial, i.e.,
\begin{equation}
\rho\left(\chi\right) \simeq \sum_{i=0}^{i_\mathrm{max}} k_i \chi^i\label{eq:alphas}
\end{equation}
Here, $i_\mathrm{max}=9$. The coefficients $k_i$ have been determined by fits to our experimentally-derived plasma rate coefficients. They are provided in Tab.~\ref{tab:PRC} along with the parameters $I_0$ which are to be used in Eqs.~\ref{eq:scaled_temp} and \ref{eq:prc}. In the temperature ranges $2\times10^5 - 3\times10^7$~K  for W$^{11+}$--W$^{17+}$ and $4\times10^5 - 8\times10^7$~K for W$^{18+}$, the parameterised rate coefficients deviate by less than 2\% from the present experimentally-derived plasma rate coefficients. In particular, these temperature ranges comprise the temperatures where W$^{11+}$--W$^{18+}$ ions are expected to form in collisionally ionised plasmas (cf.~Fig.~\ref{fig:PRC}c).  At temperatures outside these ranges the deviations increase rapidly with decreasing and increasing temperature. The total uncertainty of the fitted rate coefficients is $\pm7.3\%$ up to $T\approx 2\times10^6$~K. At higher temperatures it increases up to $\pm20.1\%$ for $T>3\times10^7$~K (figure \ref{fig:PRC}b).

\section{Conclusions and Outlook}\label{sec:Conclusions}

Absolute cross sections for EISI of W$^{q+}$ ions with $11\leq q \leq18$  have been measured for electron-ion collisions energies ranging from below the lowest ionisation threshold up to 1000~eV. In addition, theoretical cross sections have been calculated within the confines of the SCADW method. Absolute plasma rate-coefficients were derived from the experimental and scaled theoretical cross sections. The present experimentally-derived rate coefficients have been compared with the ones from the ADAS data base \cite{ADAS2017} which are based on the CADW calculations of Loch \etal \cite{Loch2005}. Significant discrepancies are found particularly at temperatures where W$^{11+}$--W$^{18+}$ ions are expected to form in collisionally ionised plasmas. These discrepancies are largely rooted in the inaccurate treatment of the EISI cross sections by the CADW method (and also by the SCADW method) in the vicinity of the ionisation threshold.

Improvements on the agreement between theory and experiment require more detailed LLDW calculations as already suggested by Loch \etal and as explicitly shown by Zhang and Kwon \cite{Zhang2014} for W$^{17+}$ as well as in a recent study of EISI of W$^{5+}$ ions \cite{Jonauskas2019a}. This latter study also includes a treatment of the metastable primary ions that were present in the W$^{5+}$ experiment \cite{Stenke1995}. Unknown fractions of metastable primary ions were also present in the W$^{11+}$ to W$^{18+}$ ion beams of the experiments discussed here. This potentially also contributes to the observed discrepancies between the present experimentally-derived plasma rate coefficients and the ones from the ADAS data base. More quantitative conclusions require extensive LLDW calculations which will be presented in follow-up studies \cite{Jin2019} for  some of the tungsten ions investigated here.

The accuracy of the present experimentally derived EISI rate coefficients is also limited by the fact that the experimental electron-ion collision energy was constrained to a maximum value of 1000~eV. In order to improve on this constraint, a new electron gun has recently been put into operation at the Giessen electron-ion crossed-beams setup which extents the accessible energy range to 3500~eV \cite{Ebinger2017}. This new electron gun would also facilitate an extension of the present study to tungsten ions with charge states higher than 19.

\section*{Acknowledgments}

Support from Deutsche Forschungsgemeinschaft (DFG) under project number Mu-1068/20, by the German Federal Ministry of Education and Research (BMBF) within the \lq\lq{}Verbundforschung\rq\rq\ funding scheme (grant no.\ 05P15RGFAA), and by the National Science Foundation of China (grant no.\ 11374365) is gratefully acknowledged. B.E. is financially supported by a grant provided within the frame of the formal cooperation between the GSI Helmholtzzentrum f\"ur Schwerionenforschung (Darmstadt, Germany) and the Justus-Liebig-Universit\"at Gie{\ss}en.

\section*{References}


\providecommand{\newblock}{}

\end{document}